\documentstyle[12pt]{article}
\begin{document}
\setlength{\oddsidemargin}{-0.1cm}
\setlength{\topmargin}{-1.cm}

\title{Separable Dual Space Gaussian Pseudo-potentials}
\author{S. Goedecker$^{(1)}$, M. Teter$^{(2)}$, J. Hutter$^{(1)}$
\\$^{(1)}$Max-Planck-Institut f\"{u}r Festk\"{o}rperforschung, Stuttgart,
Germany \\
\\$^{(2)}$Laboratory of Atomic and Solid State Physics, \\Cornell University,
 Ithaca,NY 14853-3801 \\
and Corning Inc., Corning, N.Y. 14831}
\date{}
\maketitle

\begin{center}
(Received \hspace{3cm} )
\end{center}

\vspace{1.5cm}

\noindent{\bf Abstract} \newline

We present pseudo-potential coefficients for the first two rows of
the periodic table.
The pseudo potential is of a novel analytic form, that
gives optimal efficiency in numerical calculations using plane waves
as basis set.
At most 7 coefficients are necessary to specify its analytic form.
It is separable and has optimal decay
properties in both real and Fourier space. Because of this property, the
application of the nonlocal part
of the pseudo-potential to a wave-function can be done in an efficient way
on a grid in real space. Real space integration is much
faster for large systems than ordinary multiplication in Fourier space
since it shows only quadratic scaling  with respect to the size of the system.
We systematically verify the high accuracy of these pseudo-potentials by
extensive
atomic and molecular test calculations.

\pagebreak
\section{Introduction}
Pseudo-potentials are an essential ingredient for efficient electronic
structure
calculations. First, by eliminating the core electrons the number of orbitals
that
has to be calculated is reduced. Second, the pseudo-wave-functions are much
smoother
in the core region than the all-electron wave-functions and the number of basis
functions can therefore be reduced. Especially if plane waves are used as basis
set
this reduction of the size of the basis set is essential.
The introduction of the norm-conserving property$^1$ made pseudo-potentials
an easy to handle and popular tool for electronic structure calculations.
Many attempts have since then been
made to construct norm-conserving pseudo-potentials, which
are numerically more efficient than the original ones. The introduction
of the separable form$^2$
of the norm-conserving pseudo-potentials was a major advance.
In spite of all these improvements there is still cubic
scaling with respect to the size of the system. For large
systems this part arising from the nonlocal pseudo-potential takes
most of the computer time.
It has been recognized for a long time in different contexts$^3$, that the
cubic scaling of the nonlocal
pseudo-potential part can be circumvented by doing the integration
on a grid in real space and proposals have
been made to construct pseudo-potentials with good properties for real
space integration by modifying existing pseudo-potentials of the
Kleinman-Bylander type$^4$. The Kleinman-Bylander form was initially not
intended
for real space use and therefore does not
satisfy any optimality condition for real space integration.
In contrast to previous work we therefore start out with an analytical
form, which has all of the optimality properties with respect to real space
integration build in. A small number of
parameters is then adjusted in such a way as to reflect the properties
of different atoms. In contrast to most implementations of separable
Kleinman-Bylander forms, it is thus not necessary to store the projectors
in numerical form on dense radial grids requiring very large files.
Instead the whole information on the first two rows of the periodic table
can be condensed in a small table on less than a page.
This method thus puts real space integration of the nonlocal pseudo-potential
terms for the first time on a
systematic basis. It is at the same time extremely easy to implement in
a plane wave program because all the matrix elements can be calculated
analytically. The chosen analytical form gives nevertheless enough
freedom to impose all the well established
pseudo-potential conditions and the pseudo-potential is therefore highly
accurate.

\section{Form of the pseudo potential}
The local part $V_{loc}(r)$ of this novel pseudo-potential is given by
\begin{eqnarray}
V_{loc}(r) &=& \frac{-Z_{ion}}{r} \: erf \left( \frac{r}{\sqrt{2} r_{loc} }
\right) +
   \nonumber \\  & &
    exp \left( -\frac{1}{2} (\frac{r}{r_{loc}})^2 \right)
\left( C_1 + C_2 \left( \frac{r}{r_{loc}} \right)^2 +
       C_3 \left( \frac{r}{r_{loc}} \right)^4 + C_4 \left( \frac{r}{r_{loc}}
\right)^6  \right) \:,
\label{vloc} \end{eqnarray}
where $erf$ denotes the error function. $Z_{ion}$ is the ionic charge (i.e.
charge of
the nucleus minus charge of the core electrons), and $r_{loc}$ gives the range
of the
Gaussian ionic charge distribution leading to the $erf$ potential.

The nonlocal part of the Hamiltonian $H(\vec{r},\vec{r'})$
is a sum of separable terms
\begin{eqnarray}
H(\vec{r},\vec{r'}) & = &   \sum_{i=1}^2 \sum_{m}
  Y_{s,m}(\hat{r}) \: p^{s}_{i}(r) \: h^{s}_{i}  \:
  p^{s}_{i}(r') \:  Y^*_{s,m}(\hat{r'}) \\
  & + & \sum_{m}
  Y_{p,m}(\hat{r}) \: p^{p}_{1}(r) \: h^{p}_{1}  \:
  p^{p}_{1}(r') \:  Y^*_{p,m}(\hat{r'}) \:,
\label{vsep} \end{eqnarray}
where $Y_{l,m}$ denotes a Spherical Harmonic.
The radial projectors $p^{l}_{i}(r)$ are Gaussians, where $l$ takes on
the values $0,1$ or alternatively $s,p$.
\begin{eqnarray}
p^{l}_{1}(r)  &=& \sqrt{2} \frac{r^l e^{-\frac{1}{2} (\frac{r}{r_l})^2 }}
{r_l^{l+\frac{3}{2}} \sqrt{\Gamma(l+\frac{3}{2})} } \\
p^{l}_{2}(r)  &=& \sqrt{2} \frac{r^{l+2} e^{-\frac{1}{2} (\frac{r}{r_l})^2 }}
{r_l^{l+\frac{7}{2}} \sqrt{\Gamma(l+\frac{7}{2})} }
\:, \end{eqnarray}
They are normalized such that
$$ \int_{0}^{\infty} p^l_i(r) \: p^l_i(r) r^2 dr = 1 \:,$$
where $\Gamma$ denotes the Gamma function.
The nonlocal potential tends rapidly to zero outside the core region.

The pseudo-potential can easily be transformed in Fourier space.
Calculating the matrix elements for plane waves normalized within a volume
$\Omega$,
$\frac{1}{\sqrt{\Omega}} e^{i\vec{K}\vec{r}}$
we obtain for the local part
\begin{eqnarray}
V_{loc}(K) = &-4 \pi& \frac{Z_{ion}}{\Omega} \frac{e^{-\frac{1}{2} (K
r_{loc})^2} }{K^2}
 + \sqrt{(2 \pi)^3} \: \frac{r_{loc}^3}{\Omega} e^{-\frac{1}{2} (r_{loc} K)^2}
 \\
  & & \left( C_1 + C_2 (3-(r_{loc} K)^2)  + C_3 (15-10(r_{loc} K)^2+(r_{loc}
K)^4) \right. \nonumber \\
 & & \left. + C_4 (105-105(r_{loc} K)^2+21(r_{loc} K)^4-(r_{loc} K)^6) \right)
\nonumber
\end{eqnarray}
For the nonlocal part we obtain
\begin{eqnarray}
H(\vec{K},\vec{K'}) & = &  \sum_{i=1}^2  \sum_{m}
  Y_{s,m}(\hat{K}) \: p^{s}_{i}(K) \: h^{s}_{i} \:
  p^{s}_{i}(K) \: Y^*_{s,m}(\hat{K}) \\
  &  -   &            \sum_{m}
  Y_{p,m}(\hat{K}) \: p^{p}_{1}(K) \: h^{p}_{1} \:
  p^{p}_{1}(K) \: Y^*_{p,m}(\hat{K}) \:.
\end{eqnarray}

The projectors $p^{l}_{i}(K)$ can be calculated analytically and
the result involves degenerate hyper-geometric functions. For the
relevant cases the result is
\begin{eqnarray}
p^{s}_{1}(K)  &=& \frac{1}{\sqrt{\Omega}}
4 r_s \sqrt{2 r_s} \: \pi^{\frac{5}{4}}  e^{-\frac{1}{2} (K r_s)^2}   \\
p^{s}_{2}(K)  &=& \frac{1}{\sqrt{\Omega}}
   8 r_s \sqrt{\frac{2 r_s}{15}} \: \pi^{\frac{5}{4}}
       e^{-\frac{1}{2} (K r_s)^2} (3-(K r_s)^2)  \\
p^{p}_{1}(K)  &=& \frac{1}{\sqrt{\Omega}}
   8 r_p^2 \sqrt{\frac{r_p}{3}} \: \pi^{\frac{5}{4}}
        e^{-\frac{1}{2} (K r_p)^2} K  \\
\end{eqnarray}
We see the projectors have the same form in real and Fourier
space, a Gaussian multiplied by a polynomial. As is well
known, the minimum uncertainty wave-packet is a Gaussian. This
new pseudo-potential has therefore optimal decay properties
both in real and Fourier space. Both properties are of utmost
importance for the construction of a numerically efficient pseudo-potential.
If the multiplication of the wave-function with the nonlocal
pseudo-potential arising from an atom is done on a grid in real space,
we want the nonlocal potential to be localized in a small region
around the atom. We can then restrict the real space integration to this
small region around the atom. In addition, we do not want to use a very dense
integration grid in this region, i.e. we want the nonlocal
pseudo-potential to be reasonably smooth. The first requirement is related
to the decay properties of the pseudo-potential
in real space, the second to its decay properties in Fourier space.
The optimal compromise between both requirements is a dual
space Gaussian pseudo-potential.

Even though this pseudo-potential was primarily developed for use
in combination with plane waves as basis set, it can also easily be implemented
with Gaussians as basis functions. All the matrix elements
can be calculated analytically and with O(N) scaling.

\section{Numerical method used for finding the pseudo potential parameters}
The pseudo-potential parameters were found by a least square fitting procedure.
The penalty function involved the differences of the eigenvalues and charges
within
an atomic sphere of the all-electron and pseudo atom.
These two conditions are equivalent to the
condition for a norm-conserving pseudo-potential if they are applied to the
occupied states. In addition we also included
these differences for the first two or three unoccupied states within each
angular momentum and for the lowest state of the first two unoccupied angular
momentums. In order to have well defined excited states the atom was put in
an additional parabolic confining potential. The inclusion of these excited
states
guarantees, that the energy versus logarithmic derivative curve of the pseudo
atom
reproduces the corresponding all-electron curve over a wide energy range of
typically
half a Hartree. The sets of pseudo potential parameters which we give in Table
5 and 6
typically reproduce the eigenvalues and charges of the occupied states to
within $10^{-6}$ atomic units and of the unoccupied ones to within $10^{-3}$
atomic units.
It turns out, that this fitting procedure also ensures, that additional
requirements
 that are generally considered$^{5,6}$ to lead to pseudo-potentials of very
high quality
such as extended norm-conservation and hardness are automatically satisfied as
will be
discussed later.

\section{Discussion of the pseudo-potential parameters}
We found that exactly one projector is necessary per orthogonalization
constraint.
For first row atoms there is therefore only one projector for the s-electrons,
for
second row atoms there are two for the s-channel and one for the p-channel.
For the alkaline and earth alkaline atoms (Li,Be,Na,Mg) we included the
outermost
shell of core electrons as valence electrons, since these core levels are very
shallow
in energy and extended in space.
Since the set of PSP parameters in table 1 are quasi minimal, they exhibit
the trends across the periodic table in the same way as other physically
meaningful quantities do (Fig.1).

    \begin{figure}		% produce figure
     \begin{center}
      \setlength{\unitlength}{1cm}
       \begin{picture}( 8.,6.8)           % figure dimensions
        \put(-3.,0.0){\includegraphics{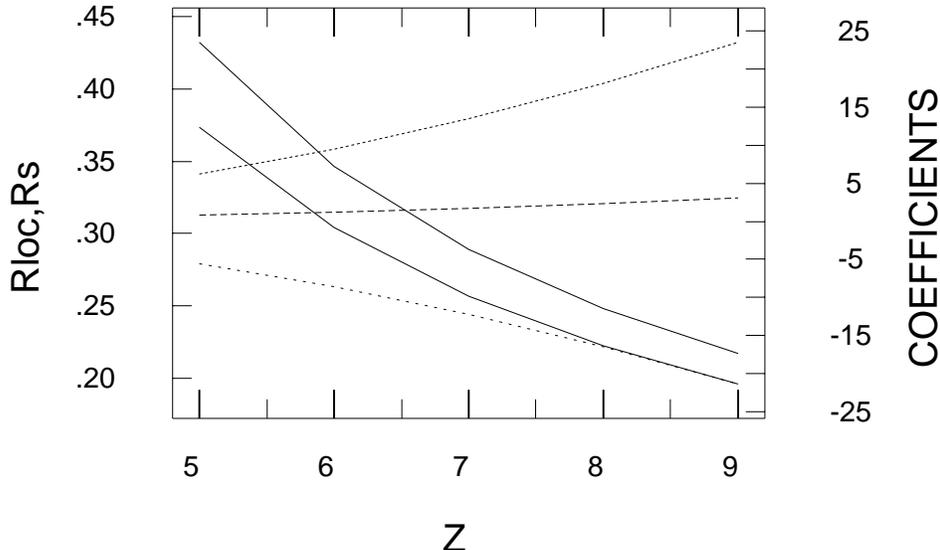}}   % VAX
       \end{picture}
\vspace{-.5cm}
       \caption{ {\it The pseudo-potential coefficients (Table 5) exhibit clear
          trends along the periodic table. On this figure the decay constants
          $r_{loc}$ and $r_s$ are shown by the solid line going with the right
          axis and the coefficients $C_1,C_2$ and $h^s_1$ by the three dotted
lines
          going with the left axis. All these parameters were found by the
least
          square fitting program. } }
\end{center}
\end{figure}

The parameters $r_{loc}, r_s$ and $r_p$ are not comparable with the parameter
$r_c$ from other pseudo-potentials. For many pseudo-potentials the
wave-function
of the pseudo and all-electron atom agree outside $r_c$. In our case they
approach
each other exponentially without ever strictly coinciding. The rate at which
they
approach is of course related to these parameters. In Fig. 2 and 3 the
wave-functions for C and Si are shown.

    \begin{figure}		% produce figure
     \begin{center}
      \setlength{\unitlength}{1cm}
       \begin{picture}( 8.,6.8)           % figure dimensions
        \put(-3.,0.0){\includegraphics{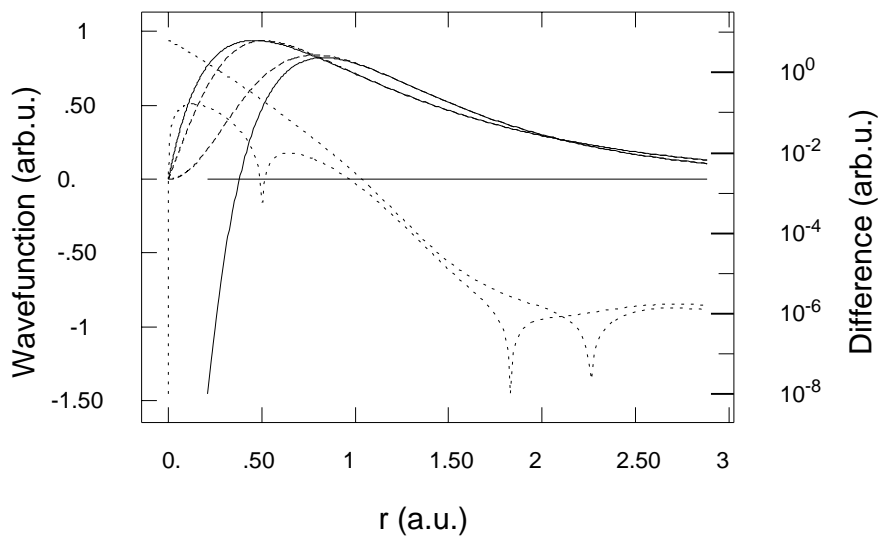}}   % VAX
       \end{picture}
\vspace{-.5cm}
       \caption{ {\it The wave-functions (full line) and pseudo-wave-functions
                  (dashed line) for Carbon. The difference between them is
                  shown by the dotted line on a logarithmic scale} }
\end{center}
\end{figure}

    \begin{figure}		% produce figure
     \begin{center}
      \setlength{\unitlength}{1cm}
       \begin{picture}( 8.,6.8)           % figure dimensions
        \put(-3.,0.0){\includegraphics{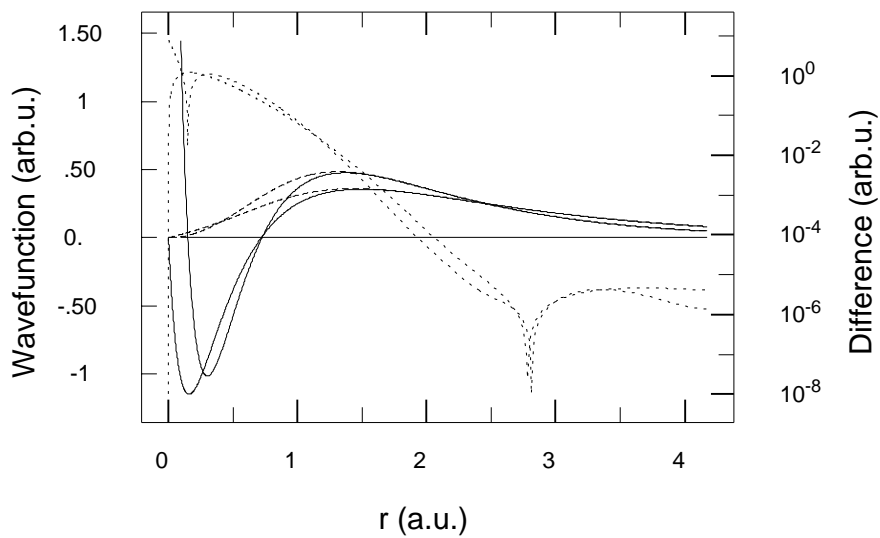}}   % VAX
       \end{picture}
\vspace{-.5cm}
       \caption{ {\it The wave-functions (full line) and pseudo-wave-functions
                  (dashed line) for Silicon. The difference between them is
                  shown by the dotted line on a logarithmic scale} }
\end{center}
\end{figure}

There are two factors, that determine the convergence of all the relevant
quantities
with respect to the plane wave energy cutoff. The first factor is the location
of the
peak of the pseudo-wave-function. In order to get a just qualitatively
reasonable
result, the the minimal wave-length representable by the plane wave basis set
has
to be equal to roughly 4 times the radius of this maximum. Once this criterium
is
satisfied systematic convergence starts. If the wave-function is analytic, the
convergence will be exponential. Because in this pseudo-potential both the
local and nonlocal potentials are analytic, the wave-function is analytic as
well and one has therefore optimal asymptotic convergence. The only thing, that
would allow to make the pseudo-potential softer would therefore be to shift
the maximum of the pseudo-wave-functions outward. This leads however to a very
fast deterioration of the physical properties of the pseudo potential.
In the construction of these pseudo-potentials we did therefore not
trade accuracy for extreme softness. We also found, that by taking a harder and
accurate pseudo-potential with a relatively low energy cutoff, one obtains
results
that are of comparable quality as the ones obtained with a softer and less
accurate
pseudo-potential at the same energy cutoff. In the second case it is just much
more difficult to realize that the results are inaccurate.
In Fig. 4 we show some examples of the convergence of the energy and
bond length with respect to the plane wave energy cutoff.
The fact, that the convergence curve is a nearly perfect straight line on
logarithmic scale shows that the asymptotic convergence sets in very early.

    \begin{figure}		% produce figure
     \begin{center}
      \setlength{\unitlength}{1cm}
       \begin{picture}( 8.,6.8)           % figure dimensions
        \put(-3.,0.0){\includegraphics{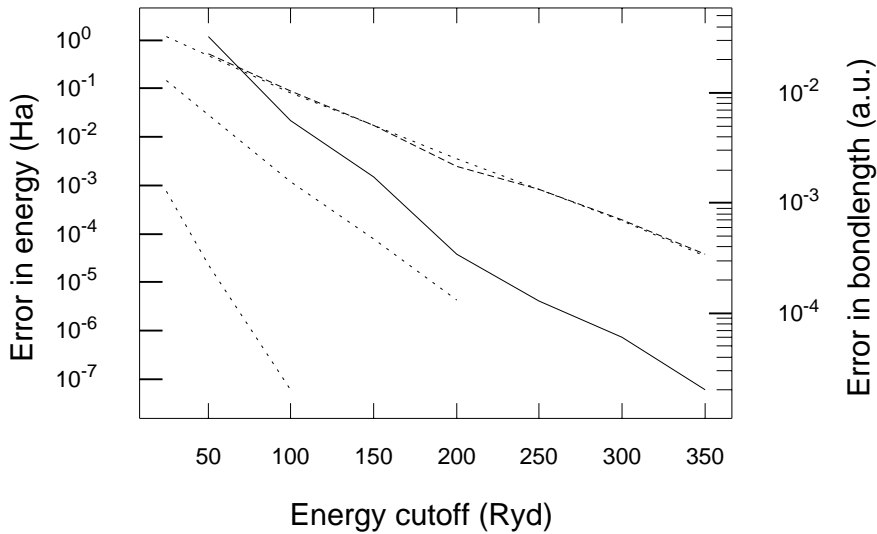}}   % VAX
       \end{picture}
\vspace{-.5cm}
       \caption{ {\it The three dotted lines show the convergence of the total
        energy for Si, C and O (in the order of increasing hardness). The
dashed
        line going with the left hand axis shows the convergence of the total
        energy for the CO molecule, the solid line going with the right axis
        the convergence of the bond length in the same molecule. Not
surprising,
        the total energy convergence in the molecule is determined and nearly
        identical to the one of the harder Oxygen.
                  } }
\end{center}
\end{figure}

As can be seen from table 5 and 6, the length scales of the pseudo-potential
are
typically a third to a forth of the covalent radius of the atom. In a real
space
implementation, the integration volume of the projectors can therefore
typically
be restricted to within a sphere whose radius is slightly larger than the
covalent
radius. This means, that the integration spheres do not appreciably overlap,
and the real space method is therefore already faster for system of very small
size.

\section{Accuracy}
We performed many atomic and molecular calculations to test the accuracy of
this
pseudo-potential. We found that the predictive power of widely used
pseudo-potential tests such as plots of
the energy vs logarithmic derivative curve and transferability tests of
excited and ionized atoms is rather limited with respect to the target
molecular calculations.
We will therefore  just mention that the pseudo-potentials of Table 5 and 6
satisfy these tests very well and give some examples for C and Si.
In table 1 we give the transferability errors for several for several excited
and ionized
states. The reference state is the neutral atom in its spherically symetrized
ground state.

\vspace{1cm}
\noindent
\footnotesize
Table 1: Transferability errors:
\newline
\noindent
\begin{tabular}{|l|c|c|c|c|} \hline \hline
Carbon  & $s^1$ $p^3$ $d^0$ & $s^2$ $p^3$ $d^0$ &
                       $s^2$ $p^1$ $d^0$ & $s^2$ $p^2$ $d^1$ \\ \hline
$\Delta \: E$ (Ha) & .3071E+00  & .4272E-01  & .3618E+00 & .3881E+00 \\ \hline
error         (Ha) & -.36E-03  & -.22E-04  &  -.31E-03  & .19E-04  \\ \hline
\hline
Silicon  & $s^1$ $p^3$ $d^0$ & $s^2$ $p^3$ $d^0$ &
                       $s^2$ $p^1$ $d^0$ & $s^2$ $p^2$ $d^1$ \\ \hline
$\Delta \: E$ (Ha) & .2509E+00  & -.5638E-02  & .2680E+00  & .1707E+00  \\
\hline
error         (Ha) & -.19E-03  & -.58E-04  &  -.22E-03  & .11E-03   \\ \hline
\hline
\end{tabular}
\vspace{1cm}
\normalsize

In Table 2 we compare the hardness$^6$ of the all-electron and pseudo atom.
The hardness is the second derivative of the total energy with respect
to the occupation numbers, $\frac{\partial^2 E}{\partial n_i \partial n_j}$,
where $n_1$ is the occupation of the s-state and $n_2$ the occupation of the
p-state. It is thus a symmetric 2 by 2 matrix.

\vspace{1cm}
\noindent
\footnotesize
Table 2: Comparison of hardness of atom and pseudo-atom
\newline
\noindent
\begin{tabular}{|c|c||c|c|} \hline \hline
\multicolumn{2}{|c|}{All-electron atom} & \multicolumn{2}{|c|}{Pseudo atom} \\
\hline \hline
\multicolumn{4}{|c|}{Carbon} \\
\hline
.4288E+00       & .4160E+00 &   .4282E+00       & .4160E+00 \\ \hline
.4160E+00       & .4040E+00 &   .4160E+00       & .4038E+00 \\ \hline
\multicolumn{4}{|c|}{Silicon} \\
\hline
.3129E+00       & .2850E+00  &   .3120E+00      & .2846E+00  \\ \hline
.2850E+0        & .2631E+00  &   .2846E+00      & .2629E+00  \\ \hline
\hline \end{tabular}
\vspace{1cm}
\normalsize

\pagebreak
The ultimate test for any pseudo-potential are molecular calculations. We
therefore
calculated the bond lengths for a large number of molecules and compared them
with the quasi exact LDA limit as given by Dickson and Becke$^7$.
The test molecules were chosen in such a way that they contain not only
single bonds, but as well multiple bonds, which typically are shorter than
single bonds and therefore more difficult to describe with a pseudo-potential.
Also molecules whose constituent atoms have large differences in
electro-negativity
were preferably chosen. In these cases, the inert region$^8$ shrinks
considerably,
since the atom with the larger electro-negativity imposes its electronic
structure
up into regions very close to the nucleus of the less electro-negative atom.
They are thus the most difficult molecules for treatment with
pseudo-potentials.
In these molecules the bond length deviate also very much from what one would
obtain by adding the covalent radii.

 The results are shown in table 3. We see, that the
errors for compounds containing the first row atoms B,C,N,O and F as well as H
are
extremely small. These errors arising from the pseudo-potential approximation
are of the
order of a few thousands of a Bohr and thus nearly
ten times smaller than the errors arising from the LDA approximation.
The errors for compounds containing the second row atoms are larger
and comparable to the LDA errors. These relatively large errors can be traced
back to the relatively shallow outermost core shells of these second row atoms.
The errors are therefore largest for Al and smallest for Cl. We could see no
improvement of the situation by adding nonlinear core correction terms$^{14}$.
The errors of molecules containing Li, Be, Na and Mg are also very small, since
the outermost core levels were included as valence states. It was however very
difficult to calculate highly precise bond lenghts for these molecules with our
plane wave program. Because these atoms are very extended, huge boxes are
necessary
in addition to high plane wave energy cutoffs resulting in many millions of
plane waves. This large basis set results in a high numerical noise
level and we give the bond length therefore in some cases only to within 2
digits.
Also Dickson and Becke quote some of these molecules only with 2 digits of
precision in the bond length. In addition we also realized that Gaussian 94
using
a 6-311G++(3df,3pd) basis set does not agree to within a few thousands of a
Bohr with
Dicksons results as is the case for the other molecules.
In the paper by Dickson and Becke there are no results
for the 2 molecules with some of the largest differences in
electro-negativity,
namely $SiF_4$ and $BF_3$. We compare them therefore with the results obtained
from
Gaussian94$^{16}$ in Table 4. Assuming that these results are correct to within
a few thousands of an Bohr, the errors are then indeed at the upper end
of the typical limit for  errors for first and second row atoms.
In summary, we can say that we obtain for molecules with only first row
atoms an accuracy  that is nearly equal to the accuracy obtained with an
all-electron
calculation using a very good Gaussian basis set. For molecules involving
second row
atoms the accuracy is clearly inferior. In all cases the accuracy is however
much
better than what is obtained with standard  Gaussian 6-31G* basis sets$^{7,9}$
or also some
other all-electron methods$^{10}$.
The molecular calculations of Table 3 and 4 were done within a Fourier space
framework. We duplicated however some calculations in real space
and obtained indistinguishable results.

\pagebreak
\footnotesize
Table 3: Comparison of the bond lengths for several small molecules as obtained
with this pseudo-potential with the quasi exact LDA result and the experimental
data.
In the last column we give the error of the LDA result compared to experiment.
The experimental bond lengths were also taken from reference 7.

\noindent
\begin{tabular}{|l|c|c|c|c|} \hline
Molecule   & PSP Bond length (a.u.) & Diff. PSP LDA &  Diff PSP EXP  & Diff LDA
, EXP \\ \hline
H2:        &      1.447 &       .001 &       .046 &      .045 \\
Li2:       &      5.099 &      -.021 &       .048 &      .069 \\
LiH:       &      3.029 &       .000 &       .014 &      .014 \\
Be2:       &      4.515 &      -.006 &      -.115 &     -.109 \\
BH:        &      2.363 &      -.010 &       .034 &      .044 \\
CH4:       &      2.072 &       .000 &       .020 &      .020 \\
C2H2(CC):  &      2.263 &      -.006 &      -.011 &     -.005 \\
C2H2(CH):  &      2.028 &      -.002 &       .023 &      .025 \\
N2:        &      2.067 &      -.001 &      -.007 &     -.006 \\
NH3:       &      1.931 &       .001 &       .019 &      .018 \\
HCN(CN):   &      2.171 &      -.003 &      -.008 &     -.005 \\
HCN(HC):   &      2.046 &       .007 &       .033 &      .026 \\
H2O:       &      1.835 &       .002 &       .026 &      .024 \\
CO:        &      2.127 &      -.002 &      -.005 &     -.003 \\
CO2:       &      2.196 &       .001 &       .004 &      .003 \\
F2:        &      2.622 &       .007 &      -.046 &     -.053 \\
FH:        &      1.764 &       .003 &       .031 &      .028 \\
CH3F(CF):  &      2.605 &       .012 &      -.007 &     -.019 \\
CH3F(CH):  &      2.081 &      -.001 &       .012 &      .013 \\
Na2:       &      5.67  &      -.00  &      -.15  &     -.15  \\
Mg2:       &      6.46  &       .01  &      -.89  &     -.90  \\
AlH:       &      3.146 &      -.023 &       .032 &      .055 \\
SiH4:      &      2.810 &      -.011 &       .015 &      .026 \\
SiO:       &      2.831 &      -.028 &      -.022 &      .006 \\
P2:        &      3.547 &      -.025 &      -.031 &     -.006 \\
PH3:       &      2.695 &      -.009 &       .024 &      .033 \\
PN:        &      2.790 &      -.017 &      -.027 &     -.010 \\
S3:        &      3.587 &      -.022 &            &           \\
H2S:       &      2.551 &      -.005 &       .027 &      .032 \\
CS:        &      2.884 &      -.012 &      -.017 &     -.005 \\
CS2:       &      2.917 &      -.010 &      -.017 &     -.007 \\
HCl:       &      2.435 &      -.004 &       .026 &      .030 \\
CH3Cl(CCl) &      3.328 &      -.001 &      -.032 &     -.031 \\
CH3Cl(CH): &      2.072 &       .000 &       .022 &      .022 \\  \hline
\end{tabular}
\normalsize

\bigskip
\noindent
\footnotesize
Table 4: Comparison of the bond lengths for two particular difficult
molecules as obtained
with this pseudo-potential and with Gaussian94 using a 6-311G++(3df,3pd)  basis
set.

\begin{tabular}{|l|c|c|} \hline
Molecule   & PSP Bond length (a.u.) & Diff to Gaussian94 LDA b.l.  \\ \hline
BF3:       &      2.477 &       .011 \\
SiF4:      &      2.926 &      -.023 \\ \hline
\end{tabular}
\normalsize

\pagebreak
\section{The parameters}
In the following we list the parameters we found. We constructed them
both for the LDA approximation and one gradient corrected scheme,
namely the BLYP$^{11}$ scheme.
Gradient corrected schemes have been examined in detail$^{9,15}$. In generally
they do no not significantly improve bond lengths but they allow to treat
hydrogen
bonding which is important in many systems containing first row atoms.
In the case of the LDA approximation, we used a
new parameterization of the results of Ceperley and Alder$^{12}$ described in
Appendix 1. Since this parameterization does not have any discontinuities
in its derivatives, it results in less numerical noise, which was very
helpful in the minimization procedure. In addition it can also be
calculated much faster numerically. Nevertheless, one can use these
pseudo-potential parameters with any other LDA parameterization, without
changing the results on a relevant scale. The entries in table 5 and 6
have the following meaning with the notation of the previous sections:

 \noindent
 \footnotesize
 \begin{tabular}{|c||c|c|c|c|} \hline \hline
 Element   &   $Z_{nuc}$   &   $Z_{ion}$  & & \\ \hline
      $r_{loc}$ &     $C_1$ &  $C_2$ & $C_3$ & $C_4$  \\
      $r_s$ & $h^s_1$  & $h^s_2$ &  &    \\
      $r_p$ & $h^p_1$ &  &  &   \\ \hline \hline
 \end{tabular}
 \normalsize

 \pagebreak
 \footnotesize
 Table 5: LDA pseudo-potential parameters \newline
\pagebreak
 \noindent
 \begin{tabular}{|c||c|c|c|c|} \hline
 H   &   1   &   1 & & \\
      .2000000 &     -4.0663326 &       .6778322 &  &   \\  \hline
 Li  &   3   &   3 & & \\
      .4000000 &    -14.0093922 &   9.5099073 &   -1.7532723 &    .0834586  \\
\hline
 Be  &   4   &   4 & & \\
      .3250000 &    -23.9909934 &   17.1717632 &   -3.3189599 &   .1650828  \\
\hline
 B   &   5   &   3 & & \\
      .4324996 &     -5.6004798 &       .8062843 &  &   \\
      .3738823 &      6.2352212 &  &  &     \\  \hline
 C   &   6   &   4 & & \\
      .3464730 &     -8.5753285 &      1.2341279 &  &   \\
      .3045228 &      9.5341929 &  &  &     \\  \hline
 N   &   7   &   5 & & \\
      .2889046 &    -12.2046419 &      1.7558249 &  &   \\
      .2569124 &     13.5228129 &  &  &     \\  \hline
 O   &   8   &   6 & & \\
      .2477535 &    -16.4822284 &      2.3701353 &  &   \\
      .2222028 &     18.1996387 &  &  &     \\  \hline
 F   &   9   &   7 & & \\
      .2168956 &    -21.4068490 &      3.0763646 &  &   \\
      .1957693 &     23.5641867 &  &  &     \\  \hline
 Na  &   11   &   9 & & \\
      .2463178 &    -22.5984025 &      3.2558639 &  &   \\
      .1688905 &     30.5987103 &  &  &     \\  \hline
 Mg  &   12   &   10 & & \\
      .2300716 &    -27.2076704 &      3.9727355 &  &   \\
      .1544802 &     36.6930557 &  &  &     \\  \hline
 Al  &   13   &   3 & & \\
      .4500000 &     -6.8340578 & & & \\
      .4654363 &      2.8140777 &      1.9395165 &  &    \\
      .5462433 &      1.9160118 &  &  &   \\  \hline
 Si  &   14   &   4 & & \\
      .4400000 &     -6.9136286 & & & \\
      .4243338 &      3.2081318 &      2.5888808 &  &    \\
      .4853587 &      2.6562230 &  &  &   \\  \hline
 P   &   15   &   5 & & \\
      .4300000 &     -6.6409658 & & & \\
      .3907376 &      3.6582627 &      3.1506638 &  &    \\
      .4408459 &      3.2859445 &  &  &   \\  \hline
 S   &   16   &   6 & & \\
      .4200000 &     -6.5960716 & & & \\
      .3626143 &      4.2228399 &      3.6696625 &  &    \\
      .4053110 &      3.8853458 &  &  &   \\  \hline
 Cl  &   17   &   7 & & \\
      .4100000 &     -6.8903645 & & & \\
      .3389943 &      4.9069762 &      4.1601818 &  &    \\
      .3762100 &      4.4850412 &  &  &   \\  \hline
 \end{tabular}
 \normalsize

 \pagebreak
 \footnotesize
 Table6: BLYP pseudo-potential parameters \newline
 \noindent
 \begin{tabular}{|c||c|c|c|c|} \hline
 H   &   1   &   1 & & \\
      .2000000 &     -4.1056068 &       .6927866 &  &   \\ \hline
 Li  &   3   &   3 & & \\
      .4000000 &    -14.1025524 &      9.6502666 &     -1.7906317 &
.0857313  \\ \hline
 Be  &   4   &   4 & & \\
      .3250000 &    -24.0585866 &     17.2528607 &     -3.3323927 &
.1653050  \\ \hline
 B   &   5   &   3 & & \\
      .4240868 &     -6.0874360 &       .9809158 &  &   \\
      .3711409 &      6.3273454 &  &  &     \\  \hline
 C   &   6   &   4 & & \\
      .3376330 &     -9.1284708 &      1.4251261 &  &   \\
      .3025277 &      9.6507303 &  &  &     \\  \hline
 N   &   7   &   5 & & \\
      .2819591 &    -12.7547870 &      1.9485936 &  &   \\
      .2554443 &     13.6593500 &  &  &     \\  \hline
 O   &   8   &   6 & & \\
      .2424499 &    -17.0170608 &      2.5613312 &  &   \\
      .2210835 &     18.3555618 &  &  &     \\  \hline
 F   &   9   &   7 & & \\
      .2128041 &    -21.9265797 &      3.2654621 &  &   \\
      .1948884 &     23.7399249 &  &  &     \\  \hline
 Na  &   11   &   9 & & \\
      .2466726 &    -22.4558069 &      3.2678153 &  &   \\
      .1687218 &     30.5372232 &  &  &     \\  \hline
 Mg  &   12   &   10 & & \\
      .2375893 &    -26.4510785 &      3.9383420 &  &   \\
      .1552995 &     36.0363418 &  &  &     \\  \hline
 Al  &   13   &   3 & & \\
      .4500000 &     -5.5482217 & & & \\
      .5058376 &      3.0200831 &      1.0641845 &  &    \\
      .5775716 &      1.5352783 &  &  &   \\  \hline
 Si  &   14   &   4 & & \\
      .4400000 &     -5.9796611 & & & \\
      .4449267 &      3.4401982 &      1.8812944 &  &    \\
      .5036368 &      2.2882053 &  &  &   \\  \hline
 P   &   15   &   5 & & \\
      .4300000 &     -5.8728328 & & & \\
      .4035454 &      3.8761979 &      2.5413108 &  &    \\
      .4527508 &      2.9405005 &  &  &   \\  \hline
 S   &   16   &   6 & & \\
      .4200000 &     -6.0083024 & & & \\
      .3704011 &      4.3736224 &      3.1957311 &  &    \\
      .4130790 &      3.5910959 &  &  &   \\  \hline
 Cl  &   17   &   7 & & \\
      .4100000 &     -6.3986998 & & & \\
      .3438408 &      4.9895061 &      3.7943315 &  &    \\
      .3813668 &      4.2346666 &  &  &   \\  \hline
 \end{tabular}
 \normalsize

\section{Conclusions}
We have presented a novel pseudo-potential. It is extremely easy to implement
both in real and Fourier space since all terms are given analytically and not
numerically. Its optimality property resulting from its dual space Gaussian
form guarantees optimal efficiency when it is used in real space. It is
highly accurate and even in the worst case of compounds containing
Al, Si and P the errors of this pseudo-potential do not dominate the
errors arising from density functional theory. In all the molecular
calculations
we did, we could not find a single molecule, where the error in the bond length
due to the pseudo-potential approximation was larger than one percent.

\section{Acknowledgments}
We thank the Cornell Theory Center for the access to its SP2, where all the
heavy computations were done. S. G. thanks Doug Allan and Pietro Ballone for
interesting discussions as well as M. Parrinello for useful remarks on the
manuscript.

\section{Appendix}
 In this work we used the following parameterization of the exchange
correlation energy functional:
 $$ \epsilon_{xc} = - \frac{ a_0+r_s (a_1+r_s (a_2+r_s a_3)) }
                       { r_s (b_1+r_s (b_2+r_s (b_3+r_s b_4))) } \: ,$$
where
\begin{eqnarray*}
     a_0 & = & .4581652932831429, \\
     a_1 & = & 2.217058676663745, \\
     a_2 & = & 0.7405551735357053,  \\
     a_3 & = & 0.01968227878617998, \\
     b_1 & = & 1.0, \\
     b_2 & = & 4.504130959426697, \\
     b_3 & = & 1.110667363742916, \\
     b_4 & = & 0.02359291751427506 \: .
\end{eqnarray*}
The exchange correlation energy in atomic units of Hartree is then given by
$$ E_{xc}= \int \rho(r) \epsilon_{xc}(\rho(r)) dr $$
where the $r_s$ parameter is related to the density $\rho$ by
$\frac{4 \pi}{3} r_s^3 = \frac{1}{\rho}$.
This parameterization was obtained by a fit to what is considered to be the
most
precise results for the exchange correlation energy$^{12,13}$. In Table 7 we
give
the best available energy as well as the differences to the widely used
parameterization
of Perdew and Zunger$^{13}$ and to our Pade approximation. As can be seen the
accuracy of both
parameterizations is very similar. However the Pade form  of the exchange
correlation
energy and its derivative (the exchange correlation potential) can be evaluated
6 times
faster (on a IBM RS6000 workstation) and it leads to less noise since it does
not
have any discontinuities in higher derivatives at $r_s$=1. Since the two
approximations
are so similar, we could not find any signficant difference in physical
observables
such as bond lengths for the two paraneterizations.

 \bigskip
 \noindent
 \footnotesize
 Table7: Comparison of the accuracy of the approximation by Perdew Zunger and
 our Pade approximation with the best available results obtained by quantum
Monte
 Carlo methods.

 \begin{tabular}{|c||c|c|c|} \hline \hline
  $r_s$  &    best (Ha)    &  Pade - best  &  PZ - best \\ \hline
   .01 &   -.46E+02 &   -.4E-01 &  -.3E-04  \\
   .10 &   -.47E+01 &   -.4E-03 &  -.4E-04  \\
   .50 &   -.99E+00 &    .8E-03 &  -.2E-04  \\
  1.00 &   -.52E+00 &   -.2E-03 &   .5E-03  \\
  2.00 &   -.27E+00 &   -.6E-03 &   .2E-03  \\
  3.00 &   -.19E+00 &   -.2E-03 &   .3E-03  \\
  4.00 &   -.15E+00 &   -.6E-04 &   .3E-03  \\
  5.00 &   -.12E+00 &   -.2E-04 &   .2E-03  \\
  6.00 &   -.10E+00 &    .1E-03 &   .3E-03  \\
 10.00 &   -.65E-01 &   -.2E-03 &  -.6E-04  \\
 20.00 &   -.34E-01 &    .1E-03 &   .2E-03  \\
 50.00 &   -.15E-01 &   -.2E-03 &  -.2E-03  \\
100.00 &   -.79E-02 &   -.1E-03 &  -.1E-03  \\ \hline
 \end{tabular}
 \normalsize

\pagebreak
\vspace{1cm}
\underline{References}
\noindent
\newline 1) D. R. Hamann, M. Schl\"{u}ter and C. Chiang,
		Phys. Rev. Lett. \underline{43}, 1494 (1980)
\newline  G. B. Bachelet, D. R. Hamann, and M. Schl\"{u}ter,
		Phys. Rev. B \underline{26}, 4199 (1982)
\newline 2) L. Kleinmann and D. M. Bylander,
		Phys. Rev. Lett. \underline{48}, 1425 (1982)
\newline    D. C. Allan and M. P. Teter,
		Phys. Rev. Lett. \underline{59}, 1136 (1987)
\newline 3) X. Gonze, P. Kaekell, and M. Scheffler,
		Phys. Rev. B \underline{41}, 12264 (1990)
\newline  D.J. Singh, H. Krakauer, C. Haas and A. Y. Liu,
		Phys. Rev. B \underline{46}, 13065 (1992)
\newline 4) R. D. King-Smith, M. C. Payne and J. S. Lin,
		Phys. Rev. B \underline{44}, 13063 (1991)
\newline 5) E. L. Shirley, D. C. Allan, R. M. Martin and J. D. Joannopoulos,
		Phys. Rev. B \underline{40}, 3652 (1989)
\newline 6)  R. G. Parr and W. Yang, "Density-Functional Theory of Atoms
             and molecules", Oxford University Press,1989
\newline  M. Teter,
		Phys. Rev. B \underline{48}, 5031 (1993)
\newline 7)  R. M. Dickson and A. D. Becke,
		J. Chem. Phys. \underline{99}, 3898 (1993)
\newline 8) S. Goedecker and K. Maschke,
		Phys. Rev. A \underline{45}, 88 (1992)
\newline 9) B. G. Johnson, P. M. W. Gill and J. A. Pople,
		J. Chem. Phys. \underline{98}, 5612 (1992)
\newline 10)  P. E. Bl\"{o}chl,
		Phys. Rev. B \underline{50}, 17953 (1994)
\newline   P. A. Serena , A. Baratoff and J. M. Soler,
		Phys. Rev. B \underline{48}, 2046 (1993)
\newline 11)  A. D. Becke,
		Phys. Rev. A \underline{38}, 3098  (1988)
\newline      C. Lee, W. Yang and R. G. Parr,
		Phys. Rev. B \underline{37}, 785  (1988)
\newline      B. Miehlich, A. Savin, H. Stoll and H. Preuss,
		Chem. Phys. Lett. \underline{157}, 200  (1988)
\newline 12)   D. M. Ceperley and B. J. Alder,
               Phys. Rev. Lett. \underline{45}, 566  (1980)
\newline 13)   J. P. Perdew and A. Zunger,
               Phys. Rev. B \underline{23}, 5048 (1981)
\newline 14)  S. G. Louie, S. F. Froyen and M. L. Cohen,
		Phys. Rev. B \underline{26}, 1738 (1982)
\newline 15)   C. J. Umrigar and X. Gonze,
		Phys. Rev. A \underline{50}, 3827 (1994)
\newline  C. Filippi, D.J. Singh and C.J. Umrigar,
                Phys. Rev. B \underline{50}, 14947 (1994)
\newline 16)
Gaussian 94, Revision B.2,
 M. J. Frisch, G. W. Trucks, H. B. Schlegel, P. M. W. Gill,
 B. G. Johnson, M. A. Robb, J. R. Cheeseman, T. Keith,
 G. A. Petersson, J. A. Montgomery, K. Raghavachari,
 M. A. Al-Laham, V. G. Zakrzewski, J. V. Ortiz, J. B. Foresman,
 C. Y. Peng, P. Y. Ayala, W. Chen, M. W. Wong, J. L. Andres,
 E. S. Replogle, R. Gomperts, R. L. Martin, D. J. Fox,
 J. S. Binkley, D. J. Defrees, J. Baker, J. P. Stewart,
 M. Head-Gordon, C. Gonzalez, and J. A. Pople,
 Gaussian, Inc., Pittsburgh PA, 1995.
\end{document}